\documentclass[12pt]{article}

\usepackage{setspace}
\usepackage{ametsoc}
\usepackage[pdftex]{graphics}
\usepackage{amsmath}    
\usepackage{amssymb}

\def\vec#1{\boldsymbol{\mathrm{#1}}}   % boldface vectors 

\let\3=\ss

\begin{document}

\title{\bf Statistics of an Unstable Barotropic Jet from a Cumulant
  Expansion}

\author{{J. B. Marston\thanks{\textit{Corresponding author address:}
      J. B. Marston, Department of Physics, Brown University,
      Providence, RI USA 02912-1843. \newline{E-mail:
        Brad\_Marston@brown.edu}}~ and E. Conover}\\ 
  Brown University, Providence, Rhode Island
\and
Tapio Schneider\\
California Institute of Technology, Pasadena, California
}

\amstitle
\singlespace

\begin{abstract}   
  Low-order equal-time statistics of a barotropic flow on a rotating
  sphere are investigated. The flow is driven by linear relaxation
  toward an unstable zonal jet. For relatively short relaxation times,
  the flow is dominated by critical-layer waves. For sufficiently long
  relaxation times, the flow is turbulent.  Statistics obtained from a
  second-order cumulant expansion are compared to those accumulated in
  direct numerical simulations, revealing the strengths and
  limitations of the expansion for different relaxation times.
\end{abstract}

\section{Introduction}
\label{intro}
Many geophysical flows are subject to the effects of planetary
rotation and to forcing and dissipation on large scales.  For example,
the kinetic energy of atmospheric macroturbulence is generated by
baroclinic instability and is then partially transferred to mean
flows, whose energy dissipation can often be represented by
large-scale dissipation. Statistically steady states of such flows can
exhibit regions of strong mixing that are clearly separated from
regions of weak or no mixing, implying that the mixing is non-ergodic
in the sense that flow states are not phase-space filling on phase
space surfaces of constant inviscid invariants such as energy and
enstrophy \citep{Shepherd87b}.  As a consequence, concepts from
equilibrium statistical mechanics, which rely on ergodicity
assumptions and can account for the statistics of two-dimensional
flows in the absence of large-scale forcing and dissipation
\citep[e.g.,][]{Miller90,Robert91,turkington01,majda06}, generally
cannot be used in developing statistical closures for such flows.

In this paper, we investigate the inhomogeneous statistics of what may be the
simplest flow subject to rotation, large-scale forcing, and
dissipation that exhibits mixing and no-mixing regions in
statistically steady states: barotropic flow on a rotating sphere
driven by linear relaxation toward an unstable zonal jet. Depending on
a single control parameter, the relaxation time, this prototype flow
exhibits behavior in the mixing region near the jet center that ranges
from critical-layer waves at short relaxation times to turbulence at
sufficiently long relaxation times. This permits systematic tests of
non-equilibrium statistical closures in flow regimes ranging from
weakly to strongly nonlinear.

We study a non-equilibrium statistical closure based on a second-order
cumulant expansion (CE) of the equal-time statistics of the flow.  The
CE is closed by constraining the third and higher cumulants to vanish,
and the resulting second-order cumulant equations are solved
numerically.  The CE is weakly nonlinear in that nonlinear eddy--eddy
interactions are assumed to vanish. We show that for short relaxation
times, the CE accurately reproduces equal-time statistics obtained by
direct numerical simulation (DNS).  For long relaxation times, the CE
does not quantitatively reproduce the DNS statistics but still
provides information, for example, on the location of the boundary
between the mixing and the no-mixing region---information that local
closures (e.g., diffusive closures) do not easily provide.

Section~\ref{model} introduces the equations of motion for the flow
and discusses their symmetries and conservation laws.
Section~\ref{DNS} describes the DNS, including the accumulation of
low-order equal-time statistics during the course of the simulation.
The CE and its associated closure approximation are outlined in section
\ref{cumulant}.  Section~\ref{compare} compares DNS and CE.
Implications of the results are discussed in section \ref{discuss}.

%%%%%% Model %%%%%%%%%%%%%%%%
\section{Barotropic jet on a rotating sphere}
\label{model}

\subsection{Equations of motion}

We study forced-dissipative barotropic flow on a sphere of radius $a$
rotating with angular velocity $\Omega$. Though not crucial for this
paper, we prefer to work on the sphere and not in the $\beta$-plane
approximation, as the sphere can support interesting phenomena not
found on the plane \citep[e.g.,][]{cho1996}.  The absolute vorticity
$q$ is given by
\begin{eqnarray}
  q &=& \zeta + f 
  \nonumber \\
  &=& \nabla^2 \psi + f
\end{eqnarray}
where $\zeta$ is the relative vorticity, $\psi$ is the stream
function, $\nabla^2$ is the Laplacian on the sphere, and
\begin{eqnarray}
  f(\phi) = 2 \Omega \sin\phi
\end{eqnarray}
is the Coriolis parameter, which varies with latitude $\phi$.  The
time evolution of the absolute vorticity is governed by the equation
of motion (EOM)
\begin{eqnarray}
  {{\partial q}\over{\partial t}} + 
  J[\psi, q] = \frac{q_{\rm jet} - q}{\tau},
  \label{EOM}
\end{eqnarray}
where 
\begin{eqnarray}
  J[\psi, q] \equiv \frac{1}{a^2 \cos(\phi)} 
  \left(\frac{\partial \psi}{\partial \lambda} 
    \frac{\partial q}{\partial \phi} -
    \frac{\partial \psi}{\partial \phi} 
    \frac{\partial q}{\partial \lambda} \right) 
  \label{Jacobian}
\end{eqnarray}
is the Jacobian on the sphere and $\lambda$ is longitude. Forcing and
dissipation are represented by the term on the right-hand-side of
Eq.~(\ref{EOM}), which linearly relaxes the absolute vorticity $q$ to
the absolute vorticity $q_{\rm jet}$ of a zonal jet on a relaxation
time $\tau$.

The zonal jet is symmetric about the equator and is characterized by
constant relative vorticities $\pm\Gamma$ on the flanks far away from
the apex and by a rounding width $\Delta \phi$ of the apex,
\begin{eqnarray}\label{q_jet}
q_{\rm jet}(\phi) = f(\phi) - \Gamma \tanh\left(\frac{\phi}{\Delta \phi}\right)\ .
\end{eqnarray}
The meridional profile (\ref{q_jet}) in our simulations is shown in
Fig. \ref{figure5} below. In the limiting zero-width case $\Delta \phi
\rightarrow 0$ of a point jet,
\begin{eqnarray} 
\zeta_{\rm jet}(\phi) \equiv q_{\rm jet}(\phi) - f(\phi) = -\Gamma~ {\rm sgn}(\phi)~ ,
\end{eqnarray}
and the jet velocity has zonal and meridional components
\begin{eqnarray}
u_{\rm jet}(\phi) &=&   \Gamma a \tan(| \phi |/2 - \pi/4), 
\nonumber \\
v_{\rm jet}(\phi) &=& 0.
\end{eqnarray}
For $\Gamma > 0$, the zonal velocity attains its most negative value $-\Gamma a$
at the equator.  

For $\Gamma > 0$, the gradient of the absolute vorticity (\ref{q_jet})
changes sign at the equator, so the jet satisfies the Rayleigh-Kuo
necessary condition for inviscid barotropic instability.
\citet{lindzen83} showed that the linear stability problem for the
barotropic point jet on a $\beta$-plane is homomorphic to the Charney
problem for baroclinic instability. The analog of the horizontal sign
change of the absolute vorticity gradient in the barotropic problem is
the vertical sign change of the generalized potential vorticity
gradient in the baroclinic problem (with the generalized potential
vorticity gradient including a singular surface contribution from the
surface potential temperature gradient). What is the meridional
coordinate in the barotropic problem corresponds to the vertical
coordinate rescaled by the Prandtl ratio in the baroclinic problem.

The analogy to the baroclinic Charney problem motivated extensive
study of the barotropic point-jet instability and its nonlinear
equilibration, with the forcing and dissipation on the right-hand side
of Eq.~(\ref{EOM}) as a barotropic analog of radiative forcing by
Newtonian relaxation and Rayleigh drag
\citep[e.g.,][]{schoeberl84,Nielsen84,Schoeberl86,shepherd88}.
Building on this body of work, here we focus on the statistically
steady states of the flow and study their dependence on the relaxation
time $\tau$. This allows us to test non-equilibrium closures
systematically in two-dimensional barotropic flows that exhibit
similar phenomena as analogous three-dimensional baroclinic flows,
with the caveat, of course, that additional degrees of freedom in
three-dimensional baroclinic flows, such as adjustment of the static
stability, make the equilibration of baroclinic instabilities
different from that of barotropic instabilities.

\subsection{Symmetries and conservation laws}

Because the jet to which the flow relaxes is symmetric about the
equator, steady-state statistics of the flow are hemispherically
symmetric. Deviations from hemispheric symmetry can be used to gauge
the degree of convergence towards statistically steady states. They
will also highlight a qualitative problem with the statistics
calculated by the CE (see section \ref{compare} below).

The EOM, Eq.~(\ref{EOM}), is invariant under a rotation of the
azimuth, $\lambda \rightarrow \lambda + \alpha$, and under an inversion of the coordinates,
\begin{eqnarray}
\phi &\rightarrow& -\phi
\nonumber \\
\lambda &\rightarrow& -\lambda
\nonumber \\
q &\rightarrow& q
\nonumber \\
q_{\rm jet} &\rightarrow& q_{\rm jet}\ .
\label{inversion}
\end{eqnarray}
Furthermore, the vorticities change sign under a north-south reflection
about the equator,
\begin{eqnarray}
\phi &\rightarrow& -\phi
\nonumber \\
\lambda &\rightarrow& \lambda
\nonumber \\
q &\rightarrow& -q
\nonumber \\
q_{\rm jet} &\rightarrow& -q_{\rm jet}~ .
\label{reflection}
\end{eqnarray}
These symmetries are reflected in the statistics discussed below.

As a consequence of the constancy of the relaxation time $\tau$,
statistically steady states satisfy two constraints that can be
obtained by integrating the EOM over the domain. Kelvin's circulation
and Kelvin's impulse of long-time averages $\langle \cdot \rangle$
in a statistically steady state are both equal to those of the jet to
which the flow relaxes,
\begin{eqnarray}
  \int \langle q(\vec{r}, t) \rangle ~ d\vec{r} &=& \int
  q_{\mathrm{jet}} ~ d\vec{r}, \label{circulation} \\
  \int \langle q(\vec{r}, t) \rangle \sin\phi ~ d\vec{r} &=& \int
  q_{\mathrm{jet}} \sin \phi ~ d\vec{r}, \label{impulse}
\end{eqnarray}
where $\vec{r} \equiv (\phi, \lambda)$ is a position vector.
Conservation of circulation (\ref{circulation}) is trivially satisfied
because vorticity integrals vanish at each moment in time,
\begin{eqnarray}
\int q(\vec{r}, t)~ d\vec{r} = \int \zeta(\vec{r}, t)~ d \vec{r} = 
\int q_{\mathrm{jet}}(\vec{r}, t)~ d\vec{r} = 0.
\label{sumRule}
\end{eqnarray}
However, conservation of impulse (\ref{impulse}), which is equivalent
to conservation of the angular momentum about the rotation axis, is
not trivial and must be respected by statistical closures.

%%%%%%%%% DNS %%%%%%%%%%%%%%%%%%%%%%%
\section{Direct numerical simulation}\label{DNS}

\subsection{Parameters and implementation}

All vorticities and their statistics can be expressed in units of
$\Omega$, but to give a sense of scale, we set the rotation period to
$2 \pi / \Omega = 1$ day.  We use Arakawa's \citeyearpar{arakawa66} energy- and
enstrophy-conserving discretization scheme for the Jacobian
on a $M \times N$ grid.  For all results reported
below, there are $M = 400$ zonal points and $N = 200$ meridional
points.  The lattice points are evenly spaced in latitude and
longitude, apart from two polar caps that eliminate the coordinate
singularities at the poles.  Each cap subtends 0.15 radians
($8.6^\circ$) in angular radius and, following Arakawa,
consists of a union of triangles radiating from the pole that match
the grid along their base; scalar fields are constrained to be constant in each cap.  
At the initial time $t=0$, we set $q =
q_{\rm jet}$ plus a small perturbation that breaks the azimuthal
symmetry and triggers the instability.  The time integration is then
carried out with a standard second-order leapfrog algorithm using a
time step of $\Delta t = 15\,\mathrm{s}$.  The accuracy of the
numerical calculation was checked, in the absence of the jet, against
exact analytic solutions that are available for special initial
conditions \citep{gates62}.  The jet parameters are fixed to be
$\Gamma = 0.6\Omega$ and $\Delta \phi = 0.05$ radians ($2.9^\circ$).
Though unphysically fast for Earth, the jet illustrates the strengths
and shortcomings of the CE. Code implementing the numerical
calculation is written in the Objective-C programming language, as its
object orientation and dynamic typing are well suited for carrying out
a comparison between DNS and the CE. 

\begin{figure}
\resizebox{16cm}{!}{\includegraphics{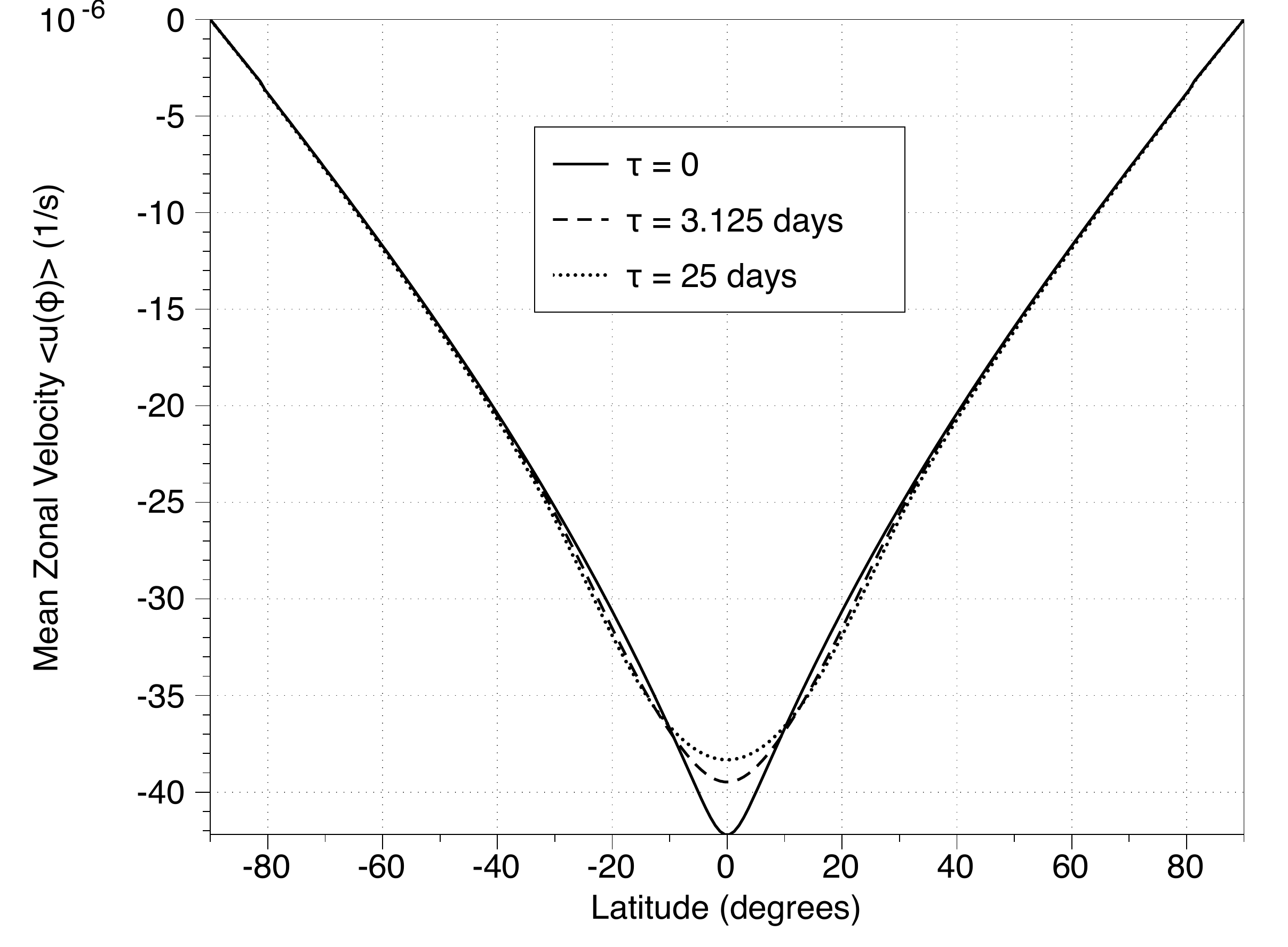}}
\caption{Mean zonal component of the velocity $\langle u(\phi)
  \rangle$ on a unit sphere of radius $a = 1$.  Relaxation times of
  $\tau = 0$ [corresponding to the profile of the fixed jet
  $u_{\mathrm{jet}}(\phi)$], $3.125$ days, and $25$ days are
  plotted.}
\label{figure1}
\end{figure}

Figure ~\ref{figure1} shows the zonal velocity of the fixed jet
$u_{\mathrm{jet}}(\phi)$ and the mean zonal velocity of the flow
$\langle u(\phi) \rangle$ for two different relaxation times.
Rounding of the jet due to mixing is evident.  The absolute vorticity
during the evolution of the instability and in the statistically
steady state eventually reached in a typical DNS are shown in
Fig.~\ref{figure2}.  Figure~\ref{figure3} displays snapshots of the
absolute vorticity in the steady-state regime for six different
choices of $\tau$.  In the limit of vanishingly short relaxation time
$\tau \rightarrow 0$ and strong coupling to the underlying jet, the
fixed jet dominates, and $q = q_{\rm jet}$ with no fluctuations in the
flow.  For $\tau > 0$, instabilities develop, and irreversible mixing
begins to occur in critical-layer waves, which form Kelvin cats' eyes
that are advected zonally with the local mean zonal flow
\citep[e.g.,][]{Stewartson81,Maslowe86}. At sufficiently large
relaxation times ($\tau \gtrsim 12$ days), the jet becomes turbulent,
and as $\tau$ increases further, turbulence increasingly homogenizes
the absolute vorticity in a mixing region in the center of the jet.
The dynamics are strongly out of equilibrium and nonlinear for
intermediate values of $\tau$, yet continue to be statistically steady
at long times.  In the limit of long relaxation time $\tau \rightarrow
\infty$ and weak coupling to the underlying jet, and upon addition of
some small viscosity to the EOM, the system reaches an equilibrium
configuration at long times
\citep{salmon98,turkington01,weichman05,majda06}, and again the
fluctuations vanish.  Here we restrict attention to the geophysically
most relevant case of short and intermediate relaxation times.

Part of what makes this flow an interesting prototype problem to test
statistical closures is that, except in the extreme limits of
vanishing or infinite relaxation time, irreversible mixing is confined
to the center of the jet and does not cover the domain. An estimate of
the extent of the mixing region can be obtained by considering the
state that would result by mixing absolute vorticity in the center of
the jet such that it is, in the mean, homogenized there and continuous
with the unmodified absolute vorticity of the underlying jet at the
boundaries of the mixing region. Because of the symmetry of the jet,
this state would have mean absolute vorticity
\begin{equation}\label{stable_q}
  \langle q \rangle = 
  \begin{cases} 
    0                & \text{for } |\phi| \le \phi_c,\\
    q_{\mathrm{jet}} & \text{for } |\phi| \ge \phi_c,
  \end{cases}
\end{equation}
and the boundaries of the mixing region would be at the latitudes at
which $q_{\mathrm{jet}} = 0$, which are, with our parameter values,
$\phi_c \approx \Gamma/(2\Omega) \approx 17^\circ$
\citep[cf.][]{schoeberl84,shepherd88}.\footnote{In the analogy to the
  Charney problem, the meridional scale $a \Gamma/(2\Omega) =
  \Gamma/\beta$ with $\beta = 2\Omega/a$ is the barotropic analog of
  the vertical \citet{held78b} scale over which quasigeostrophic
  potential vorticity fluxes associated with unstable baroclinic waves
  extend.} The meridional gradient of
the resulting mean absolute vorticity does not change sign, so the
corresponding flow would be stable according to the Rayleigh-Kuo
criterion. It represents a zonal jet that is parabolic near the
equator. However, while the mean absolute vorticity satisfies the
circulation constraint (\ref{circulation}) not only in the domain
integral but integrated over the mixing region between $\pm \phi_c$,
it does not satisfy the impulse constraint (\ref{impulse}).  To
satisfy the impulse constraint, the mixing region in a statistically
steady state extends beyond the latitudes $\phi_c$, as can be seen in
Fig.~\ref{figure3} and will be discussed further below.  Statistical
closures must account for the structure of the transition between the
mixing and no-mixing regions in this flow.

\subsection{Low-order equal-time statistics}

The first cumulant (or first moment) $c_1$ of the relative vorticity
depends only on latitude $\phi$, reflecting the azimuthal symmetry of
the EOM,
\begin{eqnarray}
c_1(\vec{r}) = \langle \zeta(\vec{r}) \rangle = c_1(\phi)~ .
\label{azimuthal1}
\end{eqnarray}
It is also convenient to define the first moment of the absolute
vorticity, $q_1(\phi) \equiv \langle q(\vec{r}) \rangle = c_1(\phi) +
f(\phi)$.  The calculation of the time averages $\langle \cdot
\rangle$ commences once the jet has reached a statistically steady
state. As the adjustment of the mean flow is controlled by the
relaxation time $\tau$, reaching a statistically steady state takes
longer for larger $\tau$. Statistics are then accumulated every $100$
minutes for a minimum of $100$ days of model time, until adequate
convergence is obtained. We have verified that the long-time averages
thus obtained are typically independent of the particular choice of
initial condition; see, for instance, Fig.~\ref{figure4}.  (The one
exception is the case of $\tau = 3.125$ days, in which, depending on
initial conditions, critical-layer waves of wavenumber three or four
are present in the statistically steady states.  As the
wavenumber-three mode has slightly lower kinetic energy, and is
reached from strongly perturbed initial conditions, we focus on it
here.)  As expected, azimuthal symmetry is recovered in such long
time-averages, as can be seen, for instance, in the final panel of
Fig.~\ref{figure2}.  In addition, the first cumulant changes sign
under reflections about the equator,
\begin{eqnarray}
c_1(-\phi) = -c_1(\phi)\ ,  
\label{c1Symmetry}
\end{eqnarray}
a consequence of the reflection symmetry (\ref{reflection}).

The second cumulant of the relative vorticity, given in terms of its
first and second moments by
\begin{eqnarray}
c_2(\vec{r}, \vec{r}^\prime) = \langle \zeta(\vec{r})~ \zeta(\vec{r}^\prime) \rangle_C \equiv \langle \zeta(\vec{r})~ \zeta(\vec{r}^\prime) \rangle - \langle \zeta(\vec{r}) \rangle \langle \zeta(\vec{r}^\prime) \rangle\ , 
\end{eqnarray}
depends on the latitude of both points $\vec{r}$ and $\vec{r}^\prime$, but only on the difference in the longitudes:
\begin{eqnarray}
c_2(\vec{r}, \vec{r}^\prime) = c_2(\phi, \phi^\prime, \lambda - \lambda^\prime)\ .
\label{azimuthal2}
\end{eqnarray}
It is essential to take advantage of the azimuthal symmetry of the
second cumulant, Eq. (\ref{azimuthal2}), to reduce the amount of memory
required to store the second cumulants by a factor of $M$, from $M^2
N^2$ to $M N^2$ scalars.  
In the DNS, the reduction is realized by averaging the second cumulant
over $\lambda^\prime$ for each value of $\Delta \lambda \equiv \lambda
- \lambda^\prime$.  The averaging also improves the accuracy of the
statistic.

By definition, the second cumulant is symmetric under an interchange of
coordinates, $c_2(\vec{r}, \vec{r}^\prime) = c_2(\vec{r}^\prime,
\vec{r})$.  It also possesses the discrete symmetry
\begin{eqnarray}
c_2(-\phi, -\phi^\prime, \Delta \lambda) = c_2(\phi, \phi^\prime, \Delta \lambda)~ , 
\label{c2Symmetry}
\end{eqnarray}
a consequence of the reflection operation (\ref{reflection}).

%%%%%% cumulant expansion %%%%%%%%%
\section{Second-order cumulant expansion}
\label{cumulant}

The jets considered here are inhomogeneous and possess nontrivial
mean flows. As a consequence, the leading-order nonlinearity is the
interaction between the mean flow and eddies (fluctuations about the
mean flow), and already the mean flow, a first-order statistic, is of
interest in closure theories \citep[e.g.][]{schoeberl84}. In contrast,
the leading-order nonlinearity in homogeneous flows is the interaction
of eddies with each other, and only higher-order statistics are of
interest in closure theories. Many higher-order closure approximations 
impose a requirement of homogeneity
\citep[e.g.,][]{holloway77,legras80,huang2001} and are not directly applicable to 
systems with inhomogeneous flows.

The second-order closure we consider is based on an expansion of
vorticity statistics in equal-time cumulants. The cumulant expansion
can be formulated either using the Hopf functional approach
\citep{frisch95,ma05} or by a standard Reynolds decomposition of each
scalar field into a mean component plus fluctuations or eddies
(denoted with a prime):
\begin{eqnarray}
q(\vec{r}) &=& \langle \zeta(\vec{r}) \rangle + f(\phi) + \zeta^\prime(\vec{r})
\nonumber \\
\psi(\vec{r}) &=& \langle \psi(\vec{r}) \rangle + \psi^\prime(\vec{r})\ .
\label{Reynolds}
\end{eqnarray}
The EOMs for the first and second cumulants
may be written most conveniently by introducing the following
auxiliary statistical quantities:
\begin{eqnarray}
p_1(\vec{r}) &\equiv& \langle \psi(\vec{r}) \rangle
\nonumber \\
p_2(\vec{r}, \vec{r}^\prime) &\equiv& \langle \psi(\vec{r}) \zeta(\vec{r}^\prime) \rangle_C
\nonumber \\
&=&  \langle \psi^\prime(\vec{r}) \zeta^\prime(\vec{r}^\prime) \rangle\ .
\end{eqnarray}
These quantities contain no new information as $c_1 = \nabla^2 p_1$
and $c_2 = \nabla^2 p_2$, where it is understood that unprimed
differential operators such as $\nabla^2$ and $J[~ ,~ ]$ act only on
the unprimed coordinates $\vec{r}$.  Substituting Eq. (\ref{Reynolds})
into Eq. (\ref{EOM}) and applying the averaging operation $\langle
\cdot \rangle$ yields the EOM for the first moment or cumulant:
\begin{eqnarray}
{{\partial c_1(\vec{r})}\over{\partial t}} &=& 
J[c_1(\vec{r}) + f(\phi),~ p_1(\vec{r})] 
+ \int J[\delta^2(\vec{r} - \vec{r}^\prime),~ p_2(\vec{r}, \vec{r}^\prime)]~ d\vec{r}^\prime
+ \frac{\zeta_{\rm jet}(\phi) - c_1(\vec{r})}{\tau}\ .
\nonumber \\
\label{1stCumulantEOM}
\end{eqnarray}
Here partial integration over $\vec{r}^\prime$ has been used to group
$\psi^\prime$ and $\zeta^\prime$ that appear as separate arguments of
the Jacobian operator into the statistical quantity $p_2$.  Similarly,
multiplication of Eq. (\ref{EOM}) by $\zeta^\prime(\vec{r}^\prime)$ followed
by averaging yields the EOM for the second cumulant, which upon
discarding the term cubic in the fluctuations may be written as
\begin{eqnarray}
{{\partial c_2(\vec{r}, \vec{r}^\prime)}\over{\partial t}} &=& 
J[c_1(\vec{r}) + f(\phi),~ p_2(\vec{r}, \vec{r}^\prime)] + 
J[c_2(\vec{r}, \vec{r}^\prime),~ p_1(\vec{r})]  
- \frac{c_2(\vec{r}, \vec{r}^\prime)}{\tau} + (\vec{r} \leftrightarrow \vec{r}^\prime),
\nonumber \\
\label{2ndCumulantEOM}
\end{eqnarray}
where $(\vec{r} \leftrightarrow \vec{r}^\prime)$ is shorthand notation
for terms that maintain the symmetry $c_2(\vec{r}, \vec{r}^\prime) =
c_2(\vec{r}^\prime, \vec{r})$.  Closure has been achieved at second
order in the expansion by constraining the third and higher cumulants
to be zero
\begin{eqnarray}
c_3 = \langle \zeta(\vec{r})~ \zeta(\vec{r}^\prime)~ \zeta(\vec{r}^{\prime \prime}) \rangle_C = 0,\  {\rm etc.}
\label{closure}
\end{eqnarray}
Otherwise an additional term would appear in
Eq.~(\ref{2ndCumulantEOM}) that couples the second and third
cumulants.  The closure approximation $c_3 = 0$ amounts to neglecting
eddy-eddy interactions while retaining eddy-mean flow interactions
\citep[e.g.,][]{herring63,schoeberl84}.

The EOMs for the two cumulants are integrated numerically using the
same algorithms and methods as those employed for DNS, starting from
the initial conditions $c_1(\vec{r}) = \zeta_{\rm jet}(\vec{r})$ and
$c_2(\vec{r}, \vec{r}^\prime) = c~ \delta^2(\vec{r} - \vec{r}^\prime)
- c / 4 \pi$ with small positive $c$.  The cumulants evolve toward the
fixed point
\begin{eqnarray}
{{\partial c_1(\vec{r})}\over{\partial t}} = 
{{\partial c_2(\vec{r}, \vec{r}^\prime)}\over{\partial t}} = 0\ .
\label{fixedPoint}
\end{eqnarray}
As a practical matter, we consider that the fixed point has been
reached when the cumulants do not change significantly with further
time evolution.  It is essential for the second cumulant to have an
initial non-zero value as otherwise it would be zero for all time,
corresponding to axisymmetric flow, which is unstable with respect to
non-axisymmetric perturbations.

The programming task of solving the equations of motion for the cumulants 
is simplified by implementing the CE as a
subclass of the DNS class, inheriting all of the lattice DNS methods
without modification.  The azimuthal symmetry of the statistics, Eqs.\
(\ref{azimuthal1}) and (\ref{azimuthal2}), and the discrete
symmetries, Eqs. (\ref{c1Symmetry}) and (\ref{c2Symmetry}), are
exploited to reduce the amount of memory required to store $c_2$ and
$p_2$.  The symmetries also speed up the calculation and help thwart
the development of numerical instabilities. The time step $\Delta t$
is permitted to adapt, increasing as the fixed point is reached.
Various consistency checks on the numerical solution are performed 
during the course of the time integration.  For instance we verify that
\begin{eqnarray}
c_2(\vec{r}, \vec{r}) = c_2(\phi, \phi, \Delta \lambda = 0) \geq 0
\end{eqnarray}
at all lattice points $\vec{r}$. Furthermore from Eq.~(\ref{sumRule})
it must be the case that
\begin{eqnarray}
\int c_1(\vec{r})~ d\vec{r} = \int c_2(\vec{r}, \vec{r}^\prime)~ d\vec{r} = 0\ .
\end{eqnarray}
Likewise, as the second-order cumulant expansion conserves Kelvin's
impulse, it follows from the impulse constraint (\ref{impulse}) that
\begin{eqnarray}
\int c_1(\vec{r}) \sin(\phi)~ d\vec{r} = \int q_{\rm jet}(\phi)
\sin(\phi)~ d\vec{r} 
\end{eqnarray}
and
\begin{eqnarray}
\int c_2(\vec{r}, \vec{r}^\prime) \sin(\phi)~ d\vec{r} = 0~ .
\end{eqnarray}
Finally the local mean kinetic energy is non-negative,
\begin{eqnarray}
\langle E(\vec{r}) \rangle = 
-p_2(\vec{r}, \vec{r}) - p_1(\vec{r}) c_1(\vec{r}) \geq 0,
\end{eqnarray}
because the statistics governed by Eqs.~(\ref{1stCumulantEOM}) and
(\ref{2ndCumulantEOM}) are realizable: They can be obtained from a
linear equation of motion for the vorticity that is driven by Gaussian
stochastic forcing \citep{orszag77,salmon98}.  The total kinetic energy
obtained by CE compares well to that determined by DNS.

%%%%%% comparison %%%%%%%%%%
\section{Comparison between DNS and CE}
\label{compare}
The equal-time statistics accumulated in the DNS can be directly
compared to the results of the CE because both calculations are based
on the same jet model with the same finite-difference approximations
on the same $M \times N$ lattice. Thus any differences between the DNS
and CE statistics may be ascribed solely to the closure approximation.
Results similar to those below are obtained on a coarser $200 \times
100$ lattice.

Figure~\ref{figure5} shows the mean absolute vorticity calculated with
the two approaches.  Closest agreement between DNS and the CE is found
at the shortest relaxation time of $\tau = 1.5626$ days.  The CE is
accurate for short relaxation times because fluctuations are
suppressed by the strong coupling to the fixed jet. The second
cumulant is reduced in size, and errors introduced by the closure
approximation that neglects the third cumulant are small.  For longer
relaxation times, the CE systematically flattens out the mean absolute
vorticity in the center of the jet too strongly. The largest absolute
discrepancy in the mean vorticity appears at an intermediate
relaxation time of $\tau = 3.125$ days.  At longer relaxation times,
the mean absolute vorticities in the DNS and CE become small in the
central jet region; however, their fractional discrepancy increases,
and the second cumulants show increasing quantitative and even
qualitative discrepancies.

Comparison of the second cumulants for $\tau = 1.5625$ days
(Fig.~\ref{figure6}) reveals a qualitative discrepancy.  The two-point
correlations as calculated in the CE exhibit wavenumber-three
periodicity, in disagreement with the wavenumber-four periodicity of
the critical-layer wave dominating the fluctuating flow component in
DNS (Fig~\ref{figure3}).  In this regard, the CE mimics the
wavenumber-three periodicity found in DNS at the longer relaxation
times.  In both DNS and CE, the correlations are
strongest in absolute value when one of the two points of the second
cumulant is located near the equator.  Interestingly, the second
cumulant from the DNS exhibits a near-exact symmetry that is not a
symmetry of the EOM,
\begin{equation}
c_2(-\phi, \phi^\prime, \Delta \lambda) \approx c_2(\phi, \phi^\prime,
\Delta \lambda)~ , 
\label{artificialSymmetry}
\end{equation}
in addition to the symmetries of Eqs. (\ref{c1Symmetry}) and
(\ref{c2Symmetry}).  This approximate symmetry, which holds exactly
for the second-order CE, may be attributed in the case of the DNS to
the small size of the third cumulant.  The fixed point of the
second-order CE as described by Eqs. (\ref{1stCumulantEOM}),
(\ref{2ndCumulantEOM}), and (\ref{fixedPoint}) possesses the
artificial symmetry, for under the north-south reflection $\phi
\rightarrow -\phi$ the Jacobian operator (\ref{Jacobian}) changes
sign, as do both $c_1(\phi)$ and $p_1(\phi)$, and the fixed point
equations remain unchanged provided that the second cumulant obeys Eq.
(\ref{artificialSymmetry}).  The artificial symmetry would, however,
be broken in general by any coupling of the second cumulant to a third
(non-zero) cumulant or, equivalently, by the inclusion of eddy-eddy
interactions, which can redistribute eddy enstrophy spatially.  Thus
the artificial symmetry (\ref{artificialSymmetry}) is an artifact of
the closure (\ref{closure}). 

Other qualitative discrepancies appear at longer relaxation times
(Fig. \ref{figure7}). For $\tau=25$ days, the second cumulant
calculated by DNS no longer shows the artificial symmetry
(\ref{artificialSymmetry}), whereas the symmetry continues to be
present in the CE due to the closure approximation. 
In contrast to the $\tau = 1.5625$ case, the largest two-point correlations occur when one
of the two points is away from the equator, reflecting the fact that
correlations are washed out by the strong turbulence near the jet
center. Finally, the second cumulant as calculated by CE shows a
wavenumber-three periodicity, with excessively strong correlations at
large separations, as a result of the neglect of eddy-eddy
interactions, which strongly distort the eddy fields in the DNS.
Nonetheless, even for relatively long relaxation times for which
differences between the CE and the DNS at the center of the jet are
apparent, the CE does capture the structure of the transition from the
mixing region in the center of the jet to the non-mixing region away
from the center, where the mean absolute vorticity in the DNS and the
absolute vorticity of the underlying jet coincide.

Figure \ref{figure8} compares the eddy diffusivity $\kappa$ of
absolute vorticity in the meridional direction as calculated by DNS
and CE for the two cases of $\tau = 3.125$ and $25$ days.  The
diffusivity is defined by second-order statistics:
\begin{align}
\kappa(\phi) &= - \langle v^\prime(\vec{r}) q^\prime(\vec{r}) \rangle
\left( \frac{\partial \langle q(\vec{r}) \rangle}{\partial (a \phi)}
\right)^{-1}
&& \text{(DNS)}
\nonumber \\
&= \frac{1}{\cos \phi} \frac{\partial p_2(\phi, \phi, \Delta \lambda)}{\partial (\Delta \lambda)}
\left( \frac{\partial q_1(\phi)}{\partial \phi} \right)^{-1}
&& \text{(CE)}.
\end{align}
The diffusivities calculated by the two methods are qualitatively
similar: diffusion is largely confined to the mixing region of the
jet; it is negative near critical layers within the mixing region
\citep[cf.][]{schoeberl84}.  However, there are significant
quantitative differences between DNS and CE, as might be expected
given that second-order statistics will generally be poorly reproduced
in a second-order closure.

%%%%%%%%%% discussion %%%%%%%%%%
\section{Discussion and conclusions}
\label{discuss}

The barotropic flows considered here attain statistically steady
states after sufficient time has passed. They are out of equilibrium
on large scales as the fixed zonal jet to which they relax is both a
source and a sink of energy. Statistical approaches that have been
developed to describe the equilibrium states of geophysical flows in
the absence of large-scale forcing and dissipation therefore are not
applicable here.  For example, approaches based on maximizing an
entropy functional subject to constraints on energy, enstrophy, and
possibly higher-order inviscid invariants
\citep{Miller90,Robert91,salmon98,turkington01,weichman05,majda06}
assume ergodic mixing and therefore would give statistical equilibrium
states with mixing throughout the domain, rather than mixing confined
to the region in the center of the jet.

One-point closures likewise are generally not adequate for the
inhomogeneous flows we considered.
For example, as one ingredient of a diffusive one-point closure one
might postulate a linear relationship between an eddy diffusivity
$\kappa(\phi)$ and rms fluctuations in the streamfunction
$\sqrt{\langle \psi^2(\vec{r}) \rangle - \langle \psi(\vec{r})
  \rangle^2}$, which has the same physical dimension
\citep{holloway80,holloway86,stammer98}.  Figure~\ref{figure8} reveals
that fluctuations in the streamfunction persist to much higher
latitudes owing to Rossby wave dynamics outside the mixing region. A
diffusive closure in which the diffusivity is a linear function of rms
streamfunction fluctuations would therefore also lead to mean states
with mixing in an overly large fraction of the domain.

We have tested the simplest non-trivial two-point closure
approximation based on a cumulant expansion of vorticity statistics,
obtained by discarding the third and higher cumulants. This
corresponds to neglecting eddy-eddy interactions while retaining
eddy-mean flow interactions.  The second-order CE is realizable, has
no adjustable parameters (such as eddy damping terms), and is not
restricted to homogeneous flows.  For short relaxation times, the
expansion reproduces the first moment fairly accurately.  For longer
relaxation times, it is quantitatively less accurate, but it still
captures the transition from a mixing region at the center of the jet
to a no-mixing region away from the center.

\begin{figure}
  \resizebox{16cm}{!}{\includegraphics{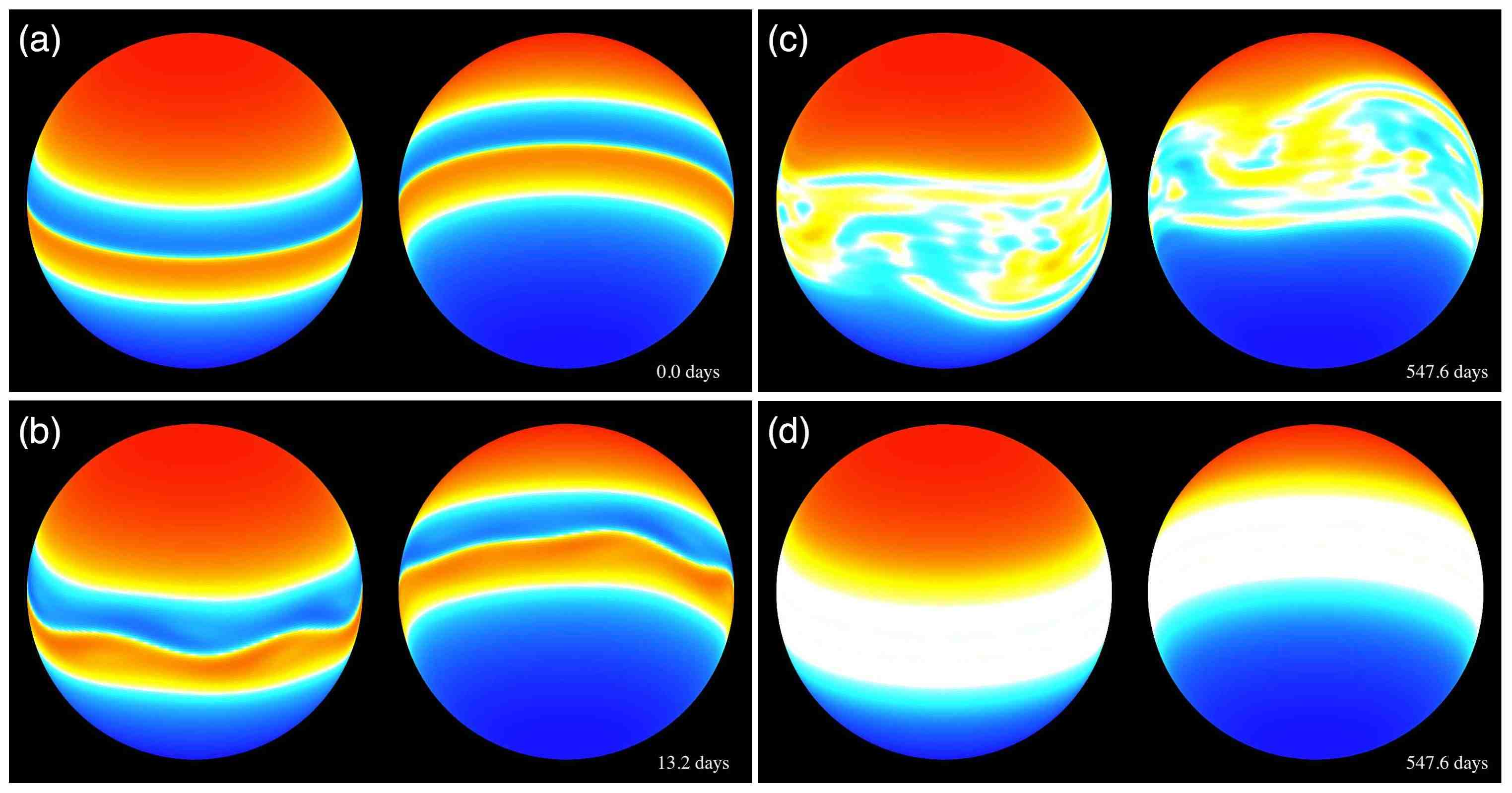}}
\caption{Absolute vorticity $q$ as calculated by DNS for a relaxation
  time of $\tau = 25$ days. The left and right hemispheres are shown
  in each panel; each is inclined by $20^\circ$ to make the poles
  visible.  Deep red (blue) corresponds to $q = \pm 10^{-4}$
  s$^{-1}$.  (a) Initial state with equatorial zonal jet.
  (b) Early development of instability. 
  (c) Statistically steady state.  (d) Mean absolute vorticity
  $q_1(\vec{r}) \equiv \langle q(\vec{r}) \rangle$ in statistically
  steady state, showing the effect of turbulence on the mean absolute
  vorticity profile and the recovery of azimuthal symmetry in the
  statistic.}
\label{figure2}
\end{figure}

\begin{figure}
\resizebox{16cm}{!}{\includegraphics{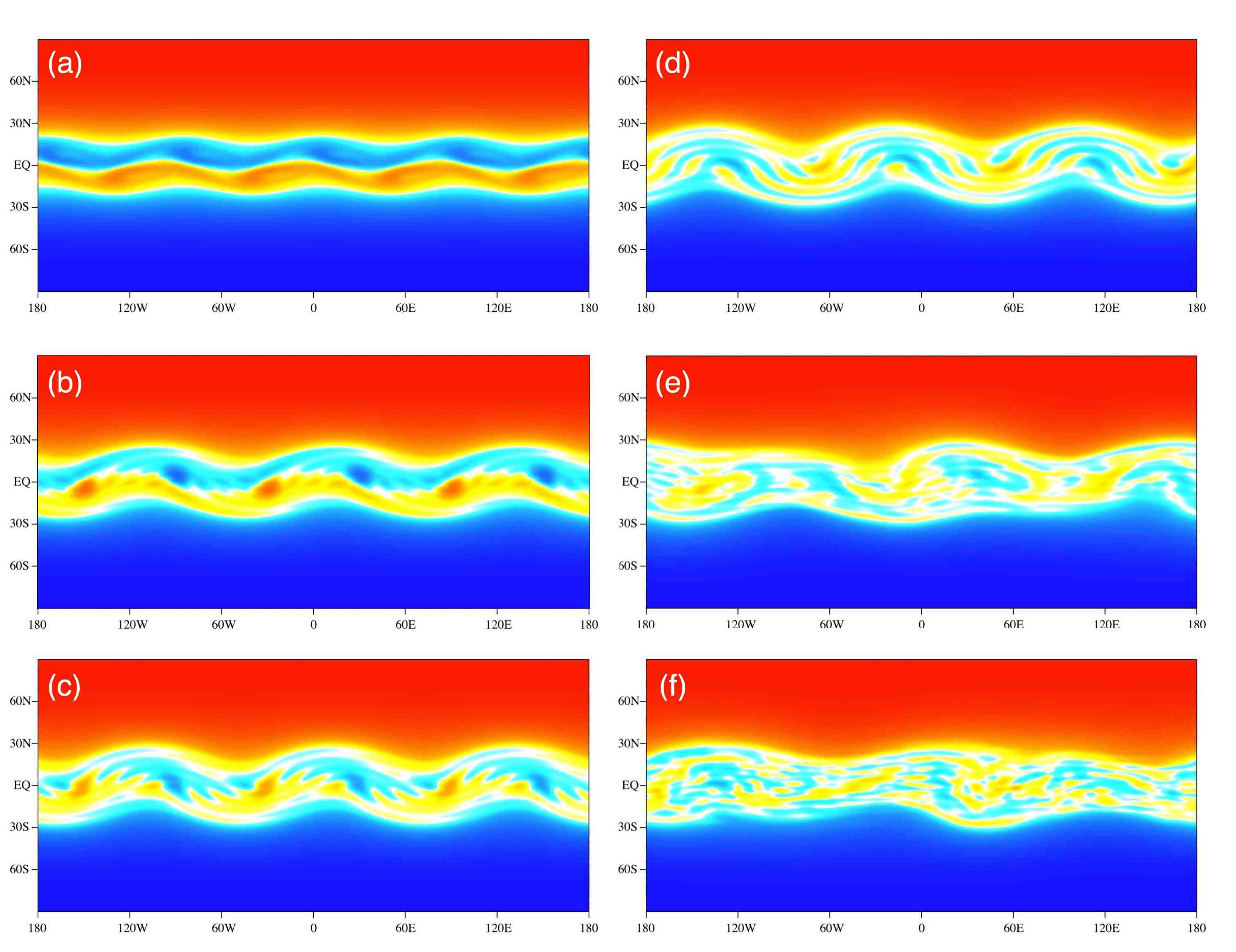}}
\caption{Snapshots of absolute vorticity in statistically steady states in a
  cylindrical projection.  The relaxation times are (a) $\tau =
  1.5625$, (b) $3.125$, (c) $6.25$, (d) $12.5$, (e) $25$, 
  and (f) $50$ days.  As in Fig. \ref{figure1}, deep red (blue) corresponds to
  $q = \pm 10^{-4}$ s$^{-1}$.}
\label{figure3}
\end{figure}

\begin{figure}
\resizebox{16cm}{!}{\includegraphics{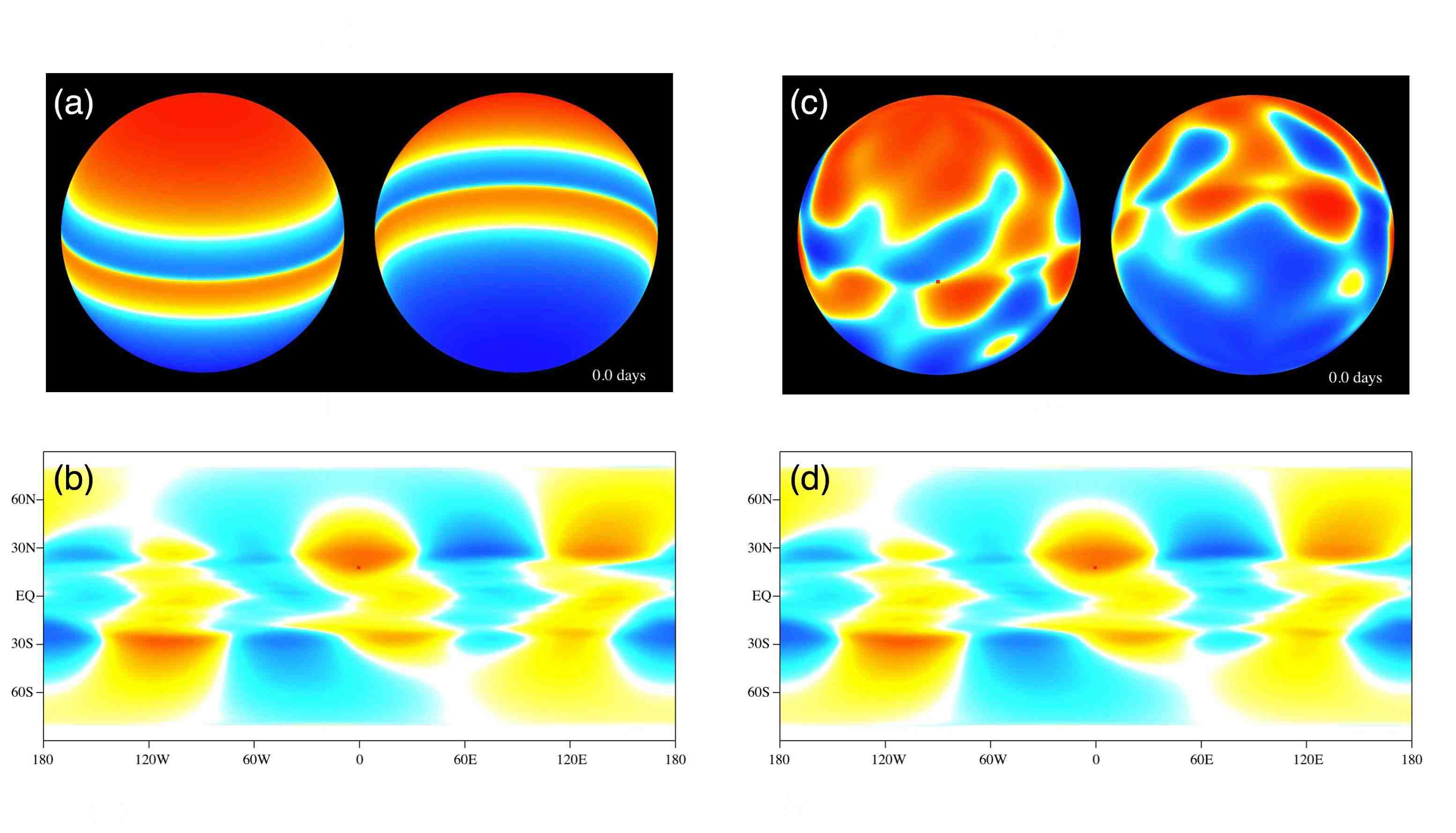}}
\vskip 0.5cm
\centerline{\resizebox{10cm}{!}{\includegraphics{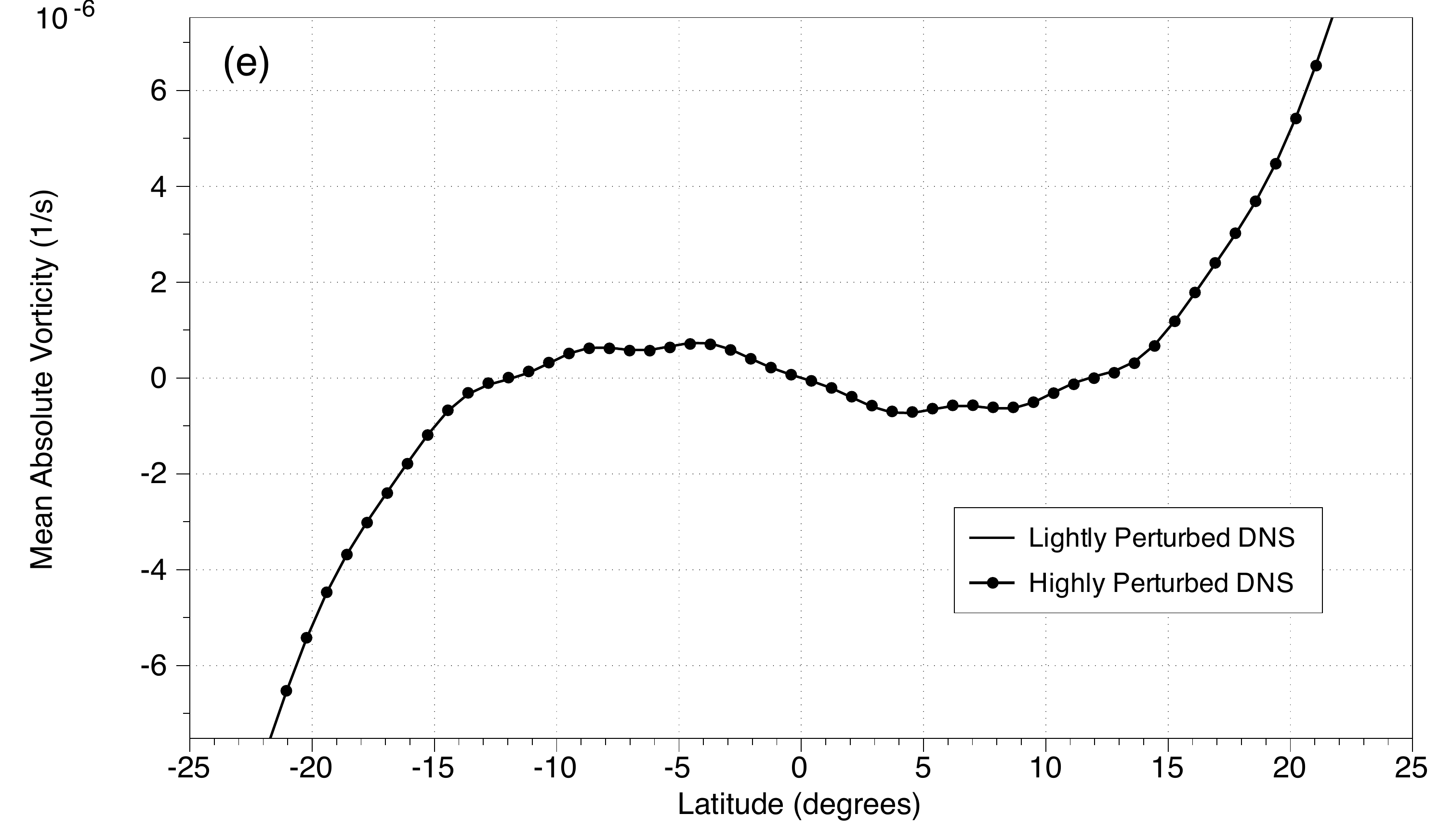}}}
\caption{Different initial conditions yield the same low-order equal
  time statistics.  The case of relaxation time $\tau = 25$ days is
  illustrated.  (a) Lightly perturbed initial absolute
  vorticity (from Fig. \ref{figure1}).  (b) Second cumulant
  obtained from the lightly perturbed initial condition with reference
  point (orange square) positioned along the central meridian ($\lambda^\prime = 0$) and at latitude $\phi^\prime = 18^\circ$.  Colors indicate positive
  (deep red is $10^{-10}$ s$^{-2}$) and negative (deep blue is
  $-10^{-10}$ s$^{-2}$) correlations with respect to the reference
  point. (c) Highly perturbed initial condition. (d) Second cumulant obtained from the highly
  perturbed initial condition. (e) Comparison of the zonally
  averaged mean absolute vorticity in the central jet region.}
\label{figure4}
\end{figure}

The steady-state statistics from the CE can be found with much less
computational effort than that required to calculate time-averaged
statistics using DNS, as the partial differential equations governing
the fixed point (\ref{fixedPoint}) are time-independent. This is
especially true if a good initial guess is available for the cumulants
$c_1$ and $c_2$ because the fixed point can then be reached rapidly by
iteration.  Furthermore, as the statistics vary much more slowly in
space than any given realization of the underlying dynamics (see Fig.
\ref{figure2}), it may be possible to employ coarser grids without
sacrificing accuracy.  Thus the CE realizes a program envisioned by
\citet{lorenz67} long ago by solving directly for the statistics, but
it does so at the cost of a closure approximation that compromises the
accuracy of the statistics, especially for flows with more strongly
nonlinear eddy-eddy interactions. There is evidence, however, that
nonlinear eddy-eddy interactions in Earth's atmosphere are
weak \citep{schneider2006}. \citet{ogorman2007} have shown
that several features of atmospheric flows, such as scales of jets and
the shape of the atmospheric turbulent kinetic energy spectrum, can be
recovered in an idealized general circulation model in which eddy-eddy
interactions are suppressed.  In particular, baroclinic jets form spontaneously 
and are maintained by eddy-mean flow interactions even in the absence 
of nonlinear eddy-eddy interactions.  So a second-order CE may be worth
exploring for more realistic models.  Extensions of the CE to
multi-layer models governed by either quasigeostrophic dynamics or the
primitive equations are straightforward.  The storage requirement for
the second cumulant grows as the square of the number of layers, but
it remains feasible so long as the models retain azimuthal symmetry.
The incorporation of topography and other symmetry-breaking effects is
more problematic because of the much larger storage requirement.

Whether more sophisticated closures can be devised that are more
accurate and yet only require comparable computational effort remains
an open question.  In the case of isotropic turbulence,
renormalization-group inspired closures show some promise
\citep{mccomb2004}, but these typically make extensive use of
translation invariance in actual calculations.  Investigation of more
sophisticated approximations for non-isotropic and inhomogeneous
systems, such as the barotropic flows we considered, may be warranted
in view of the partial success of the cumulant expansion
reported here.

\begin{acknowledgment}
  We thank Greg Holloway, Paul Kushner, Ookie Ma, and Peter Weichman
  for helpful discussions.  This work was supported in part by the
  National Science Foundation under grants DMR-0213818 and
  DMR-0605619.  It was initiated during the Summer 2005 Aspen Center
  for Physics workshop ``Novel Approaches to Climate,'' and J.~B.~M.\
  and T.~S.\ thank the Center for its support.
\end{acknowledgment}
\clearpage

\bibliographystyle{ametsoc}
\bibliography{bibliography/references}

\clearpage

\begin{figure}
  \resizebox{15cm}{!}{\includegraphics{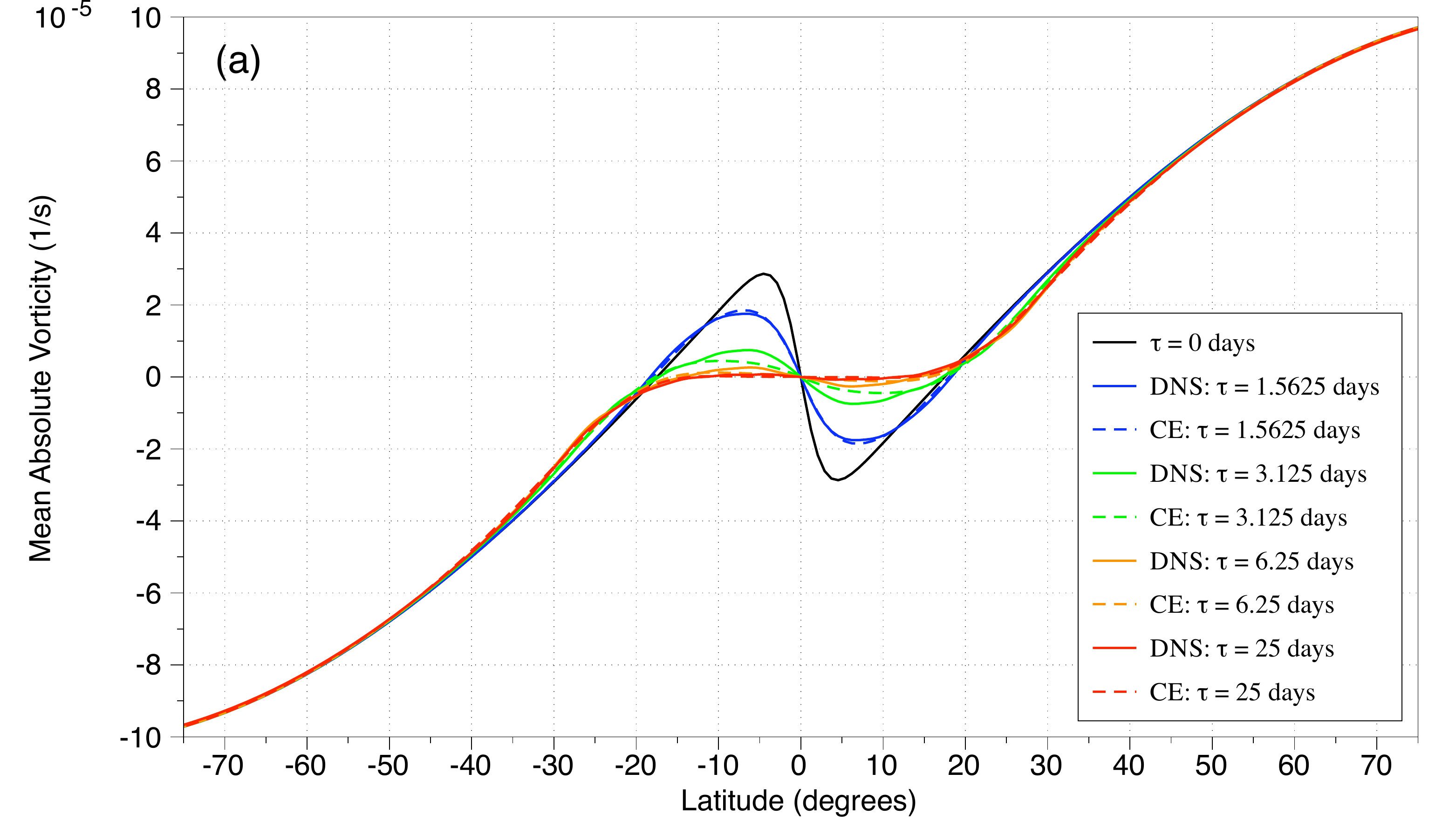}}
  \resizebox{15cm}{!}{\includegraphics{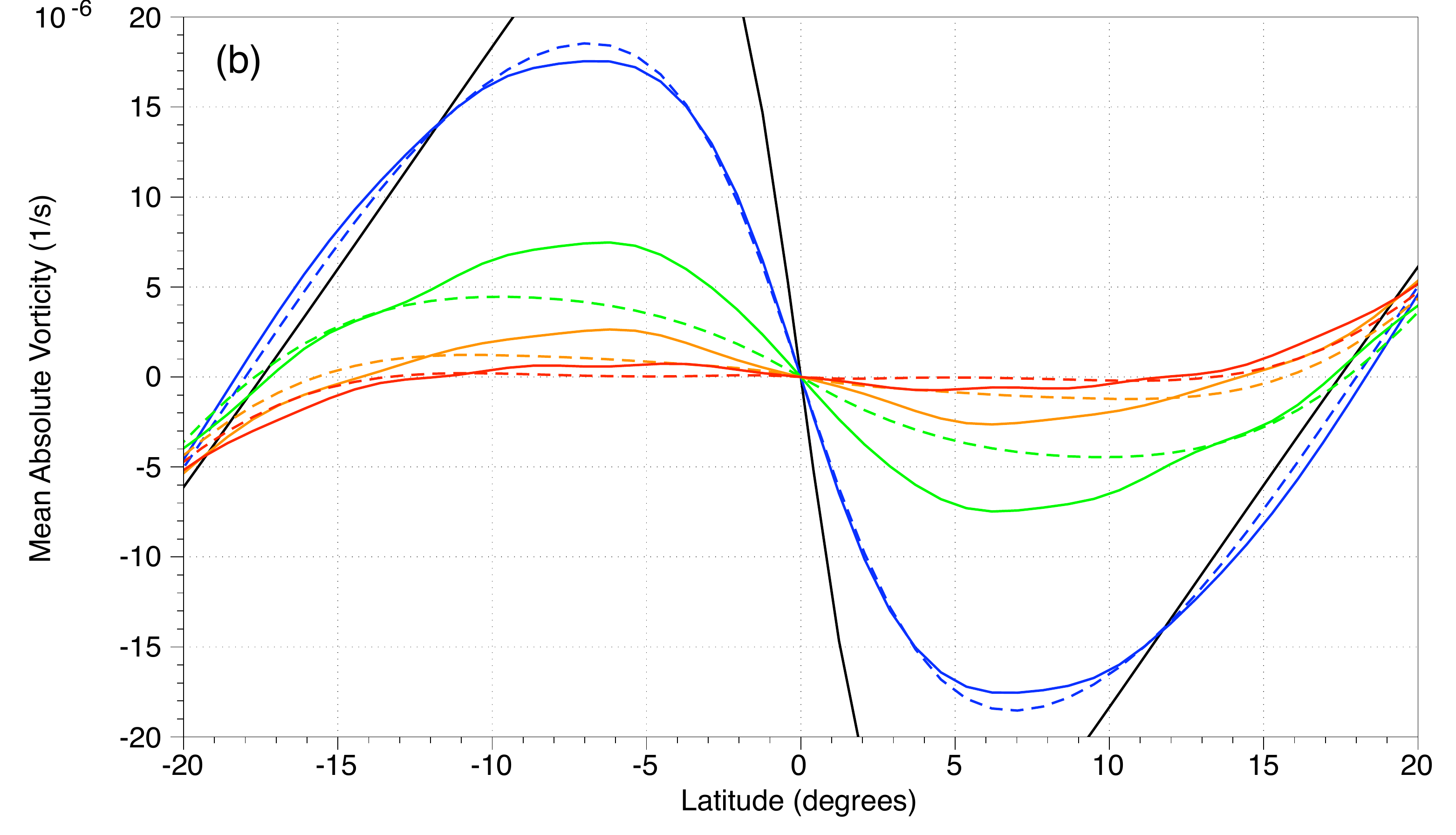}}
\caption{(a) Mean absolute vorticity $q_1$ as a
  function of latitude for different relaxation times.  Zonally averaged results from
  DNS (solid lines) are compared to those from the CE (dashed lines).
  The black line ($\tau = 0$) is the absolute vorticity of the fixed jet
  $q_{\rm jet}(\phi)$.  (b) Magnified view of central jet region.
  Note the antisymmetry of the mean absolute vorticity (the first
  cumulant) under equatorial reflections.}
\label{figure5}
\end{figure}

\begin{figure}
\resizebox{16cm}{!}{\includegraphics{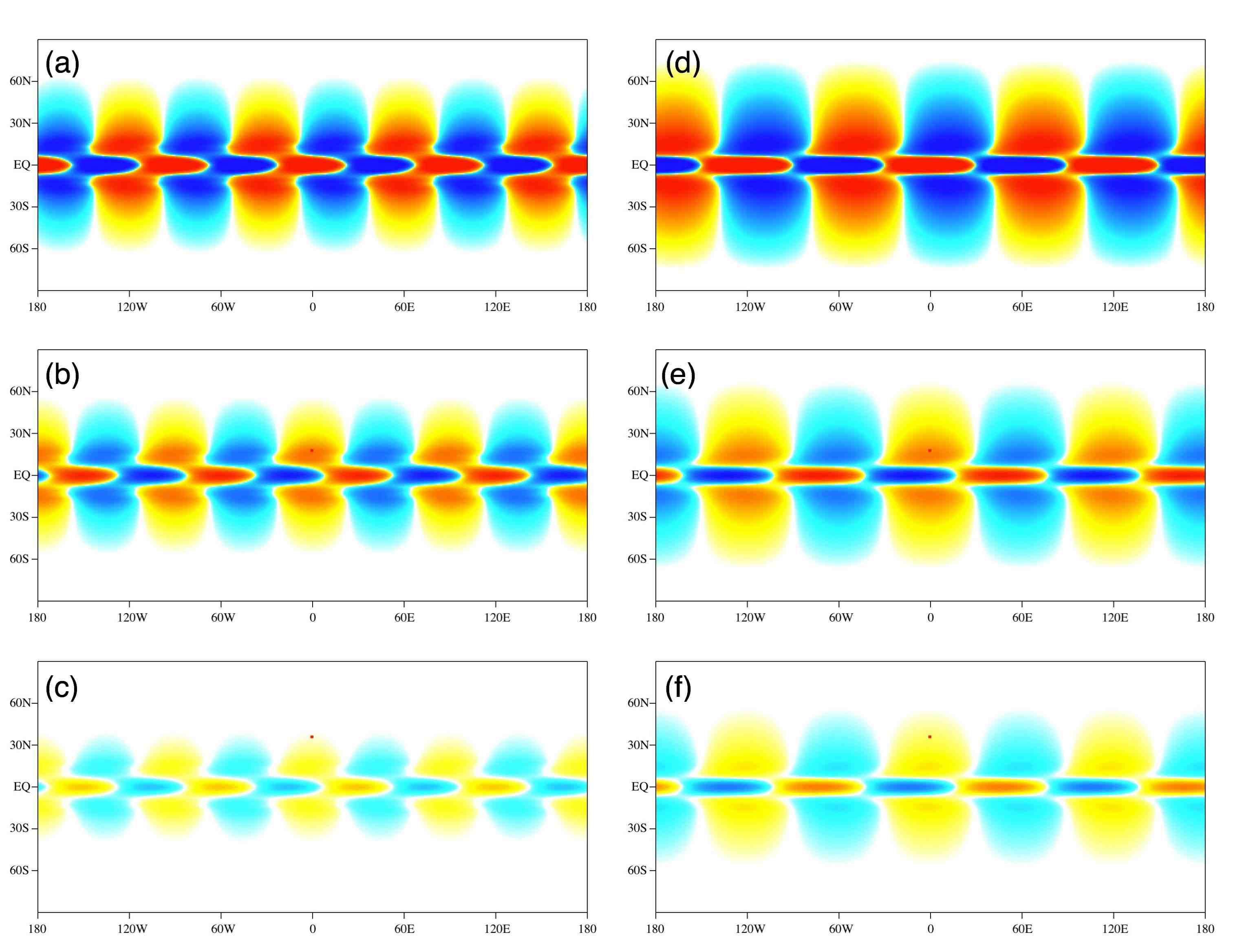}}
\caption{The second cumulant of the relative vorticity field,
  $c_2(\phi, \phi^\prime, \lambda-\lambda^\prime)$, for 
  relaxation time $\tau = 1.5625$ days.  (a), (b) and (c): DNS.  (d), (e), and (f): CE.  
  The reference point (orange square) is positioned along
  the central meridian ($\lambda^\prime = 0$) and at latitudes of $\phi^\prime = 0$ for (a) and (d),
  $\phi^\prime = 18^\circ$ for (b) and (e), and $\phi^\prime =
  36^\circ$ for (c) and (f).  Colors indicate positive (deep red is
  $10^{-10}$ s$^{-2}$) and negative (deep blue is $-10^{-10}$
  s$^{-2}$) correlations with respect to the reference point.}
\label{figure6}
\end{figure}

\begin{figure}
\resizebox{16cm}{!}{\includegraphics{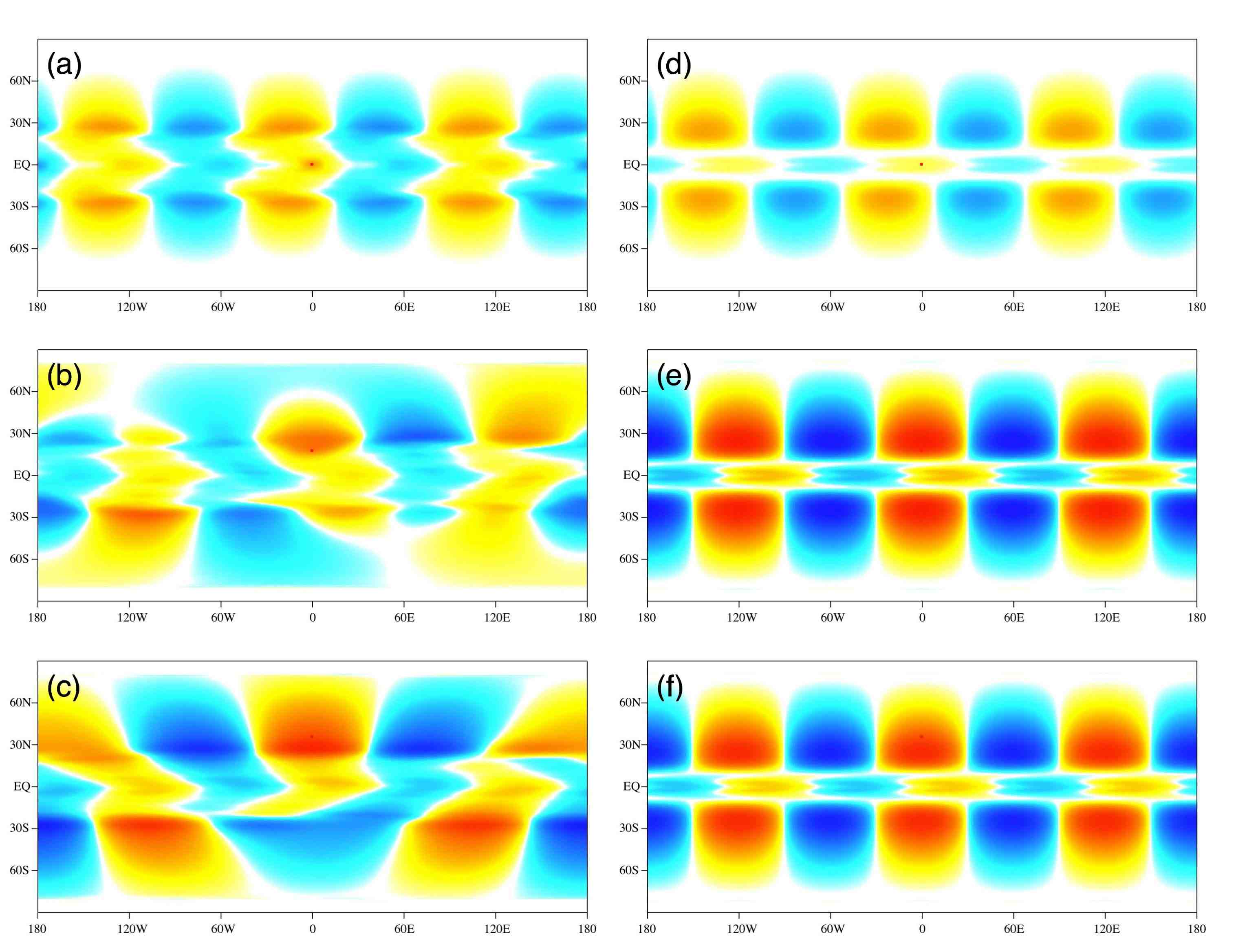}}
  \caption{Same as Fig.~\ref{figure5} except for a relaxation time of
    $\tau = 25$ days.  The reflection symmetry about the equator seen
    in the CE, an artifact of the closure truncation, is not present
    in the DNS.}  \label{figure7}
\end{figure}

\begin{figure}
  \resizebox{15cm}{!}{\includegraphics{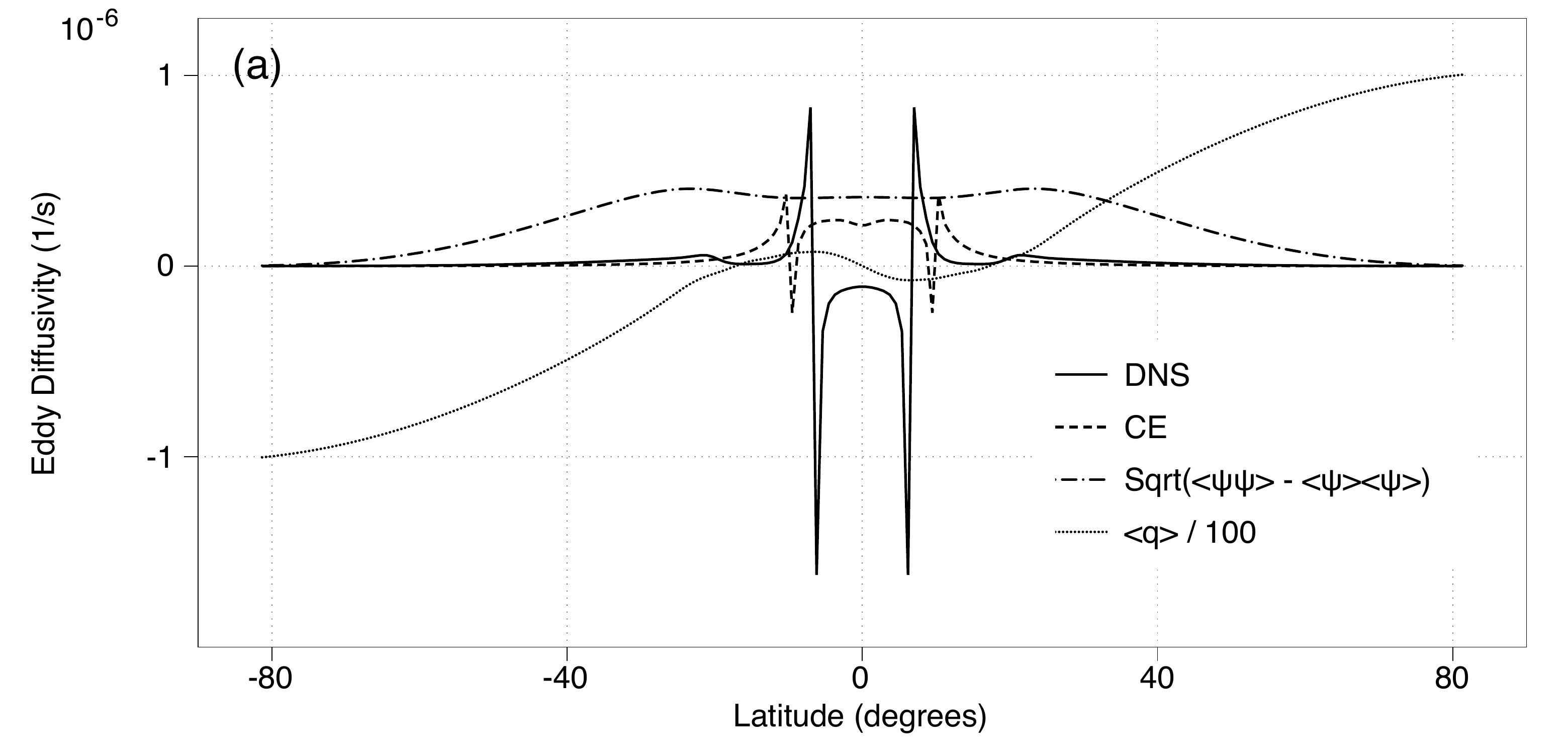}}
  \resizebox{15cm}{!}{\includegraphics{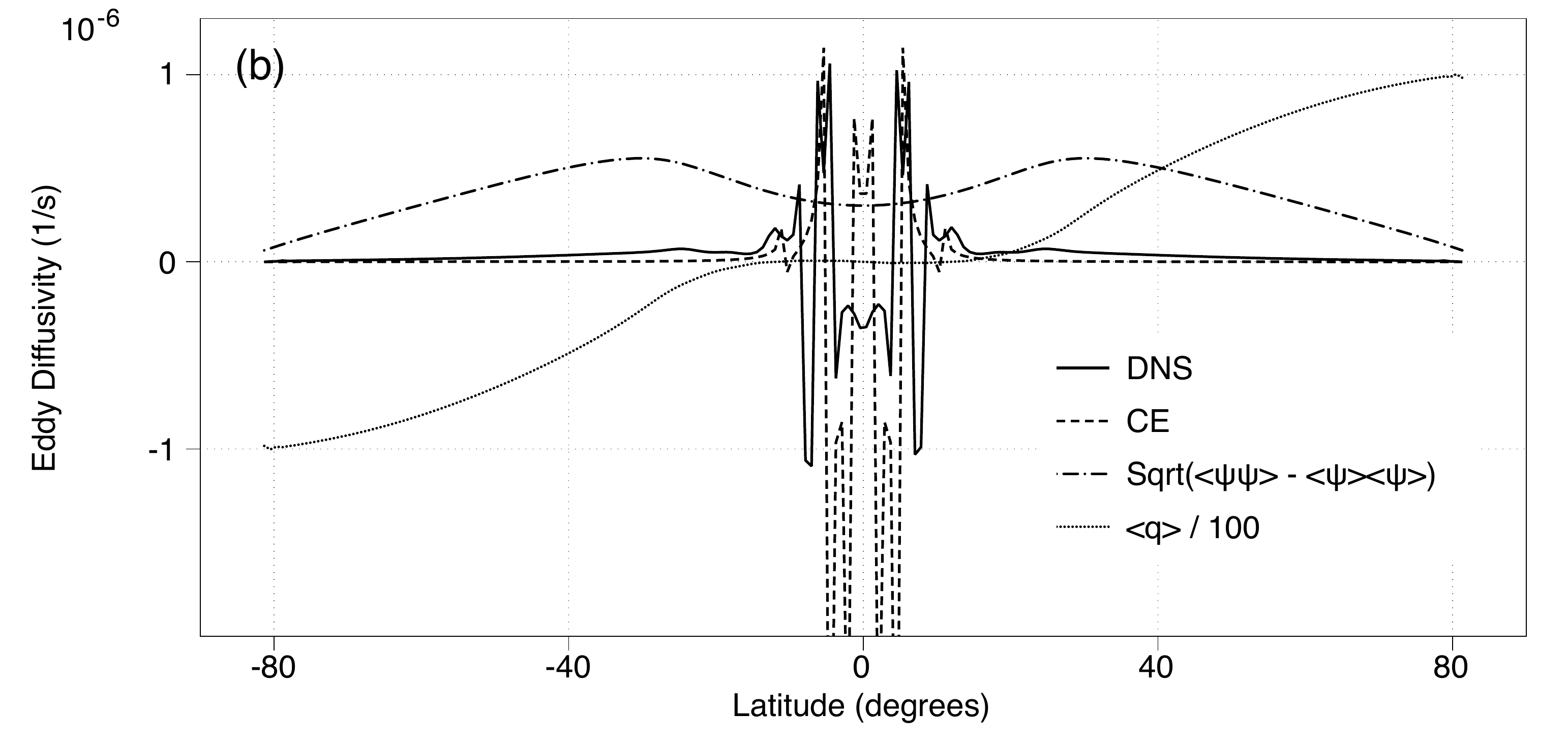}}
\caption{(a) Eddy diffusivity $\kappa$ for the case of relaxation time $\tau = 3.125$ days on the unit sphere ($a = 1$) as calculated by DNS and CE.  Also plotted for comparison are the DNS calculation of the square root of the variance in the streamfunction, $\sqrt{\langle \psi^2(\vec{r}) \rangle - \langle \psi(\vec{r}) \rangle^2}$, and the first moment of the absolute vorticity.  (b) Same as (a) except for $\tau = 25$ days.}
\label{figure8}
\end{figure}

\end{document}